\documentclass[shortnote,twocolumn]{jpsj3}
\usepackage{txfonts}
\usepackage{color}

\title
{
The magnetization process of the spin-one triangular-lattice Heisenberg
antiferromagnet
}

\author
{
J. Richter\thanks{E-mail address:
Johannes.Richter@Physik.Uni-Magdeburg.DE}, O. G\"otze, R. Zinke,
D. J. J. Farnell$^1$, and \\
H. Tanaka$^2$}

\inst
{
Institut f\"ur Theoretische Physik, Otto-von-Guericke Universit\"at
Magdeburg, P.O.B. 4120, D-39016 Magdeburg, Germany\\
$^{1}$Division of Mathematics, Faculty of Advanced Technology,
University of Glamorgan, Pontypridd CF37 1DL, Wales, UK\\
$^{2}$Department of Physics, Tokyo Institute of Technology, Meguro-ku, Tokyo
152-8551
}

\kword{triangular-lattice antiferromagnet, magnetization
plateau,  Ba$_3$NiSb$_2$O$_9$}

\begin{document}
\maketitle

% \section{Introduction}
Starting with Wannier's \cite{wannier} and Anderson's \cite{And} seminal papers the triangular-lattice
antiferromagnet has
attracted a lot of attention.
One spectacular feature of this model system for a strongly frustrated
magnet is     
the unconventional  magnetization process of the triangular-lattice
Heisenberg antiferromagnet.\cite{nishi1986,chub1991,Hon1999,ono2003,HSR04,lauchli2006,alicea2009,
fortune2009,farnell2009,motrunich2010,zhito2011,sakai2011,ono2011,takano2011,shirata2012} 
While for the classical isotropic Heisenberg model at zero temperature the  magnetization
$M$
increases linearly with the applied magnetic field $H$, thermal or quantum
fluctuations induce a plateau at 1/3 of the saturation magnetization
$M_{\rm sat}$ (`order from disorder' phenomenon).
For the extreme quantum case, i.e. spin quantum number $s\,{=}\,1/2$, this plateau at $T\,{=}\,0$ has been
widely discussed.
Very recently an almost perfect experimental realization of an $s\,{=}\,1/2$ triangular-lattice Heisenberg
antiferromagnet has been reported for Ba$_3$CoSb$_2$O$_9$\cite{shirata2012},
where the entire measured magnetization curve $M(H)$ including the 1/3 plateau is
in excellent agreement with theoretical predictions. In particular, the
theoretical magnetization data obtained by high-order coupled-cluster
approximation\cite{farnell2009} and by large-scale exact diagonalization
approach\cite{sakai2011} almost coincide with the measured ones over
a wide range of the magnetic field.

Another interesting  triangular-lattice magnet is Ba$_3$NiSb$_2$O$_9$ that is considered as a good candidate of an $s\,{=}\,1$ triangular-lattice Heisenberg
antiferromagnet.\cite{Doi,ono2011,Cheng} In Ba$_3$NiSb$_2$O$_9$, the uniform triangular lattice of magnetic ions is realized as in Ba$_3$CoSb$_2$O$_9$. Indeed, recently it has been found that
the magnetization curve of this
compound also exhibits a clear 1/3 plateau.\cite{ono2011}
By contrast to the well-investigated spin-half case there are much less
reliable theoretical studies of the $s\,{=}\,1$ model relevant for
Ba$_3$NiSb$_2$O$_9$.
Motivated by the excellent agreement of theoretical predictions based on exact
diagonalization (ED) and coupled-cluster method  (CCM) and experimental results
for the spin-half compound Ba$_3$CoSb$_2$O$_9$ reported in Ref.~\citen{shirata2012}  
we apply in this paper these methods to the $s\,{=}\,1$ model  to provide
theoretical data to compare with experimental results.

For the $s\,{=}\,1$ model the Lanczos ED  for $N\,{=}\,27$ sites and the CCM-SUB$n$-$n$
approximation\cite{subn-n} for
$n\,{=}\,2, 4, 6, 8$ are used to calculate
the field-dependent properties of the model.
In this short note we will not explain details of the Lanczos ED and the CCM. We refer  the interested reader
e.g. to Refs.~\citen{sakai2011, Poilb2004, ED40} and
\citen{farnell2009,zeng98,farnell2002,bishop04},
respectively. 
\begin{figure}
\begin{center}
\scalebox{0.42}{
\includegraphics{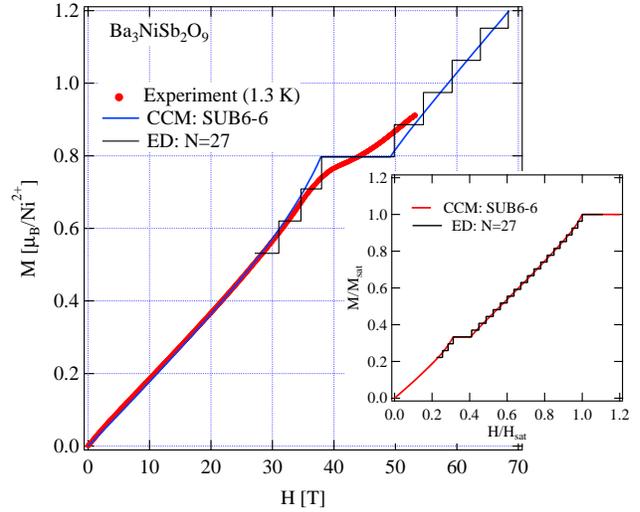}
}
\end{center}
\caption{The magnetization curve  of the 
$s\,{=}\,1$ triangular-lattice Heisenberg
antiferromagnet calculated by the CCM and by ED compared with the
experimental data for Ba$_3$NiSB$_2$O$_9$ measured at $T\,{=}\,1.3$ K, 
see Ref.~\citen{ono2011}. For the comparison we have plotted the theoretical data
using the exchange parameter $J/k_{\rm B}\,{=}\,21.9$ K and the $g$ factor
$g=2.392$. The inset shows magnetization curves up to the saturation calculated by the CCM and by ED.}
\label{fig1}
\end{figure}

\begin{figure}
\begin{center}
\scalebox{0.43}{
\includegraphics{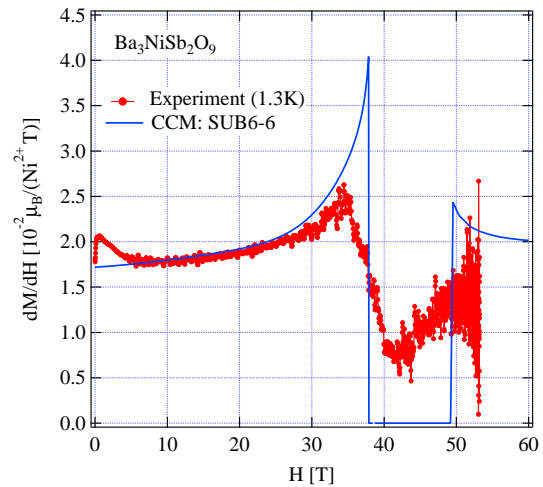}
}
\end{center}
\caption{The derivative susceptibility $dM/dH$ 
of the
$s\,{=}\,1$ triangular-lattice Heisenberg
antiferromagnet calculated by the CCM compared with the
experimental data for Ba$_3$NiSB$_2$O$_9$ measured at $T\,{=}\,1.3$ K, see 
Ref.~\citen{ono2011}. For the comparison we have plotted the theoretical data
using the exchange parameter $J/k_{\rm B}\,{=}\,21.9$ K and the $g$ factor
$g=2.392$.}
\label{fig2}
\end{figure}

\begin{figure}
\begin{center}
\scalebox{0.32}{
\includegraphics{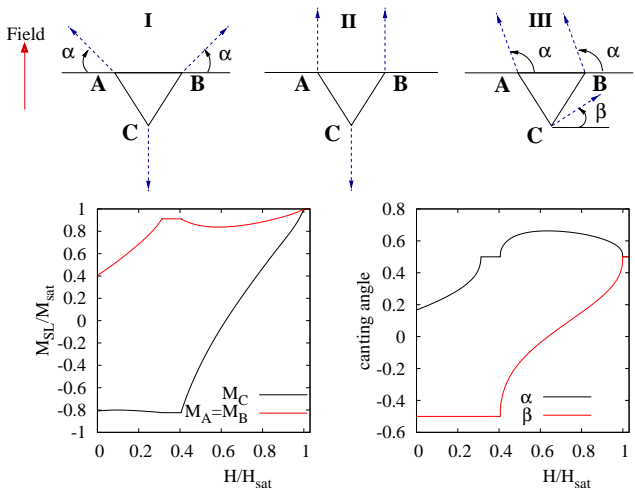}
}
\end{center}
\caption{Upper panel: Spin orderings in magnetic field. 
Lower panel: CCM-SUB6-6 data for the sublattice magnetizations $M_{\rm A,B}(H)$ and
$M_{\rm C}(H)$
(left) and the canting angles $\alpha(H)/\pi$ and $\beta(H)/\pi$ (right) as
functions
of the magnetic field $H$.}
\label{fig3}
\end{figure}

First we report the main theoretical results. The saturation field is $H_{\rm sat}\,{=}\,9Js\,{=}\,9J$,
where $J$ is the nearest-neighbor exchange coupling. 
The zero-field uniform susceptibility $\chi$ calculated with  CCM-SUB$n$-$n$
approximation is $\chi\,{=}\,0.10790$, 
$0.09932,$ $0.09785$, and $0.09679$ for $n\,{=}\,2$, 4, 6, and 8, respectively.
Moreover, it is useful 
to extrapolate the ``raw'' SUB$n$-$n$ data to $n\,{\to}\,\infty$ by
$\chi(n)\,{=}\,c_0 + c_1 (1/n) + c_2(1/n)^2$ which yields finally our CCM estimate
for the susceptibility $\chi_{\rm CCM}\,{=}\,0.0956$. This value is in excellent
agreement with the spin-wave result\cite{chub1994}, $\chi_{\rm SW}\,{=}\,0.0953$.
(We remark that this value of $\chi_{\rm S}$ in
Ref.~\citen{chub1994} was referred to as $\chi_\perp$  in this article and
furthermore
that it was defined per volume.\cite{defin_chi})
The left and right endpoints of the 1/3-plateau, $H_{\rm c1}$ and $H_{\rm c2}$ are
determined to $H_{\rm c1}/J\,{=}\,2.817 $ and $H_{\rm c2}/J\,{=}\,3.695 $ (ED, $N\,{=}\,27$),
 $H_{\rm c1}/J\,{=}\,2.727$, 2.788, 2.809, 2.8142 and $H_{\rm c2}/J\,{=}\,3.617$, 3.674, 3.648,
3.637 (CCM, SUB$n$-$n$ for $n\,{=}\,2$, 4, 6, 8, respectively).  As for the
susceptibility we may extrapolate 
to $n\,{\to}\,\infty$ 
yielding  $H_{\rm c1}/J\,{=}\,2.839$ and $H_{\rm c2}/J\,{=}\,3.552$. We can compare these values with
corresponding ones 
obtained by 
linear spin-wave theory\cite{chub1991}, $H_{\rm c1}/J\,{=}\,2.748$ and
$H_{\rm c2}/J\,{=}\,3.645$. Improving the linear spin-wave
approximation  by using a self-consistent approach Takano {\it et al.}\cite{takano2011}  
obtained $H_{\rm c1}/J\,{=}\,2.760$ and
$H_{\rm c2}/J\,{=}\,3.585$, which are closer to our CCM estimates.
The sublattice magnetization in the collinear plateau `{\it up-up-down}' state (spin
configuration II in Fig.~\ref{fig3}) calculated within CCM-SUB$n$-$n$ approximation are 
$M_{\rm up}\,{=}\,M_{\rm A,B}\,{=}\,0.93251$, 0.91565, 0.91188, 0.91097 and
$M_{\rm down}\,{=}\,M_{\rm C}\,{=}\,{-\,0.86502}$, $-$\,0.83129, $-$\,0.82375,
$-$\,0.82193  for $n\,{=}\,2, 4, 6, 8$, respectively.
Again we may extrapolate to $n\,{\to}\,\infty$ 
yielding  $M_{\rm up}\,{=}\,0.909$ and $M_{\rm down}\,{=}\,{-}\,0.817$.

Similar to Ref.~\citen{shirata2012} we now compare directly the
theoretical predictions for the ground state with low-temperature experimental results reported in
Ref.~\citen{ono2011}, see Figs.~\ref{fig1} and \ref{fig2}, where experimental data were corrected 
for the Van Vleck paramagnetic susceptibility 
of ${\chi}_{\rm VV}\approx 2.4\,{\times}\,10^{-4}\,{\rm emu/mol}=4.3\,{\times}\,10^{-4}\,{\mu}_{\rm B}/({\rm Ni^{2+}\,T})$, 
which has not been done in Ref.~\citen{ono2011}. 
Due to the finite-temperature effect, small single-ion anisotropy and small anisotropy of the $g$ factor
the experimental curve is smeared out around the endpoints of the 1/3-plateau,
$H_{\rm c1}$ and $H_{\rm c2}$. Nonetheless, the excellent agreement between theory
and experiment is obvious.
A good agreement between experimental and theoretical data is also obtained
for  the  derivative susceptibility $dM/dH$, shown in Fig.~\ref{fig2}.    
We may conclude, that the presented data demonstrate that
Ba$_3$NiSb$_2$O$_9$9 is well described by the $s\,{=}\,1$ triangular-lattice Heisenberg
antiferromagnet.

In addition to the total magnetization our CCM approach allows to calculate
the sublattice magnetizations  as well as the canting angles (see
Fig.~\ref{fig3}). The three different ground-state regimes illustrated in
the upper panel of Fig.~\ref{fig3} are clearly seen in the behavior of these quantities.
While the total magnetization $M\,{=}\,2M_{\rm up}\,{+}\,M_{\rm down}$ increases monotonously with
magnetic field  (except within the plateau region), both sublattice magnetizations,
$M_{\rm up}$ and $M_{\rm down}$, exhibit regions of negative slope $dM_{\rm SL}/dH\,{<}\,0$.

\begin{acknowledgment}

%\acknowledgment
This work was supported by a Grant-in-Aid for Scientific Research from the Japan Society for the Promotion of Science. 

\end{acknowledgment}

\end{document}